\documentclass[aps,11pt,onecolumn,preprintnumbers,amsmath,amssymb]{revtex4}

\usepackage[english]{babel}
\usepackage{graphicx}
\usepackage{color}

\begin{document}

\newcommand{\unit}[1]{\:\mathrm{#1}}            
\newcommand{\To}{\mathrm{T_0}}
\newcommand{\Tp}{\mathrm{T_+}}
\newcommand{\Tm}{\mathrm{T_-}}
\newcommand{\EST}{E_{\mathrm{ST}}}
\newcommand{\Rp}{\mathrm{R_{+}}}
\newcommand{\Rm}{\mathrm{R_{-}}}
\newcommand{\Rpp}{\mathrm{R_{++}}}
\newcommand{\Rmm}{\mathrm{R_{--}}}
\newcommand{\ddensity}[2]{\rho_{#1\,#2,#1\,#2}} 
\newcommand{\ket}[1]{\left| #1 \right>} 
\newcommand{\bra}[1]{\left< #1 \right|} 

\bibliographystyle{naturemag}

\title{Optical control of a single spin-valley in charged WSe$_2$ quantum dots}

\author{Xin Lu$^{1*,\dagger}$}
\author{Xiaotong Chen$^{1*}$}
\author{Sudipta Dubey$^{1*}$}
\author{Qiang Yao$^1$}
\author{Xingzhi Wang$^2$}
\author{Qihua Xiong$^{2,3}$}
\author{Ajit Srivastava$^{1\dagger}$}
\affiliation{$^1$Department of Physics, Emory University, Atlanta 30322, Georgia, USA.}
\affiliation{$^2$Division of Physics and Applied Physics, School of Physical and Mathematical Sciences, Nanyang Technological University, Singapore 637371, Singapore.}
\affiliation{$^3$NOVITAS, Nanoelectronics Centre of Excellence, School of Electrical and Electronic Engineering, Nanyang Technological University, Singapore, 639798, Singapore.}

\maketitle

*These authors contributed equally to this work.

$^\dagger$ Correspondence to: ajit.srivastava@emory.edu, xin.lu2@emory.edu
\textbf{Control and manipulation of single charges and their internal degrees of freedom, such as spins, is a fundamental goal of nanoscience with promising technological applications. Recently, atomically thin semiconductors such as WSe$_2$ have emerged as a platform for valleytronics, offering rich possibilities for optical, magnetic and electrical control of the valley index~\cite{XiaoPRL2012,XuNPhys2014}. While progress has been made in controlling valley index of ensemble of charge carriers~\cite{MakNNano2012,ZengNNano2012,CaoNComm2012}, valley control of individual charges, crucial for valleytronics, remains unexplored. Here, we provide unambiguous evidence for localized holes with net spin in optically active WSe$_2$ quantum dots (QDs) and control their spin-valley state with the helicity of the excitation laser under small magnetic field. We estimate a lower bound on the valley lifetime of a single charge in QD from recombination time to be $\sim$ nanoseconds. Remarkably, neutral QDs do not exhibit such a control, demonstrating the role of excess charge in prolonging the valley lifetime. Our work extends the field of 2D valleytronics to the level of single spin-valley, relevant for quantum information and sensing applications.}

Localized single spins in solid-state have been widely studied for quantum information technology, spintronics and quantum sensing~\cite{ImamogluPRL1999, GiovannettiNP2011}, in addition to serving as a versatile playground for exploring many-body physics~\cite{LattaNature2011}. With the rise of semiconducting van der Waals (vdW) materials having direct band gap such as group VI-B transition metal dichalcogenides (TMDs), a family of 2D materials for optical control of charge carriers has emerged~\cite{SplendianiNanoLett2010,MakPRL2010}. Charge carriers in TMDs possess a valley index, which is locked to their spin in presence of large spin-orbit coupling (SOC) into a spin-valley index. In contrast to spin or valley, spin-valley requires flipping of both indices for its relaxation and should remain protected in absence of short-range, magnetic impurities, making TMDs ideally suited for spin-valleytronics~\cite{XiaoPRL2012,XuNPhys2014,MakNP2016}. A distinguishing aspect of valleytronics with TMDs is that the valley index has been shown to be addressable by helicity of optical excitation. Indeed, optically generated valley polarization and coherence have been demonstrated in photoluminescence (PL) studies on MoS$_2$ and WSe$_2$, taking a step forward towards valleytronics~\cite{MakNNano2012,ZengNNano2012,CaoNComm2012,JonesNNano2013,HaoNPhys2016}. However, as recombination lifetimes of photogenerated excitations in TMDs is on the order of a few picoseconds, optically generated valley lifetime is limited to a similar timescale. On the other hand, valley polarization of free charge carriers, as opposed to photogenerated excitations, shows promising prospect with lifetimes on the order of microseconds reported for holes~\cite{JiangNComm2018,YangNPhys2015,DeyPRL2017,KimScience2017,YanPRB2017}. A natural question for quantum information science and quantum metrology applications is whether a single spin-valley can be optically addressed and manipulated. As a first step to address this question, localization of electrons or holes, for example, in quantum dots (QDs) or impurity potentials is needed. Recently discovered optically active QDs in WSe$_2$, which have orders of magnitude longer PL lifetimes ($\sim$ ns) compared to their 2D host~\cite{SrivastavaNNano2015,KoperskiNNano2015,ChakrabortyNNano2015,HeNNano2015,TonndorfOptica2015}, are an ideal candidate for this task provided that they inherit valley index from their 2D host after localization~\cite{WuPRB2016} and possess a net charge.

Here, we show that spin-valley of a single, localized hole in monolayer WSe$_2$ can be controlled by the helicity of the excitation laser. Our experiments performed on an ambipolar WSe$_2$ field effect transistor (FET) device, provide spectroscopic signatures in the emission spectra of singly charged QD comprised of two holes and an electron, which is absent in the neutral state of the same QD. Furthermore, as discussed below, due to distinctive features in the band structure of WSe$_2$, the valley of the excess hole can only be opposite to that of the electron-hole (e-h) excitation and thus can be initialized in a given state by controlling the valley index of the latter. We demonstrate this initialization by choosing the helicity of the excitation laser which results in emission corresponding to either one of the two-valley states of the excess hole, which are made spectrally distinguishable at small magnetic fields. This leads us to conclude that valley lifetime of the resident hole is at least as long as the recombination lifetime, which is on the order of a few nanoseconds. Such an optical control is absent in neutral QDs because of spin-valley mixing resulting from e-h exchange which is quenched in positive, singly charged dots. Our results lend strong support to the idea that QDs in WSe$_2$ inherit desirable valley index from the 2D host and pave the way for valleytronics on a single particle level.

QDs in WSe$_2$ are believed to be excitons trapped in shallow potential wells arising from either defects or localized strain on the monolayer flake~\cite{BrannyNComm2017,PalaciosBerraqueroNComm2017}. Furthermore, they seem to inherit the valley physics of the 2D exciton as suggested by their extreme anisotropic response with respect to an in-plane versus out-of-plane magnetic field~\cite{SrivastavaNPhys2015}. As valley mixing seem to be absent in these QDs, the length scale of confinement must be larger than the Bohr radius of exciton and trion ($\sim$ 1-2 nm)~\cite{StierNComm2016}. Thus, we can safely assume that the single-particle band structure which is used to understand the 2D exciton should be applicable to the neutral and charged QDs as well. Fig.~1a-b shows the contrast between valley configurations of the single-particle states that constitute a negatively charged QD ($X^-_{\mathrm{d}}$) and a positively charged QD ($X^+_{\mathrm{d}}$), respectively. $X^-_{\mathrm{d}}$ has two inequivalent configurations where the excess electron is in the same or opposite valley compared to the e-h pair. For a fixed spin-valley of the excess electron, the same valley and opposite valley $X^-_{\mathrm{d}}$ configurations are mixed due to e-h exchange of the exciton ($J_{eh}$) and also split by e-h exchange between the excess electron and the hole ($J'_{eh}$)~\cite{YuNComm2014}. As a result, the valleys of the exciton in $X^-_{\mathrm{d}}$ are mixed, even though the spins of the excess electron are not, leading to a loss of helicity control of the electron spin-valley (see Supplementary Information). On the other hand, the excess hole in $X^+_{\mathrm{d}}$ can only be in the opposite valley configuration, due to the large SOC in the valence band, in which case the e-h exchange interaction between the excess hole and the electron ($J'_{eh}$) is quenched. Thus, valley index of the exciton in $X^+_{\mathrm{d}}$ is perfectly anticorrelated with that of the excess hole. Moreover, $J_{eh}$ should also be strongly suppressed upon localization into a QD due to Pauli blockade (Fig. 1b). The quenching of e-h exchange in $X^+_{\mathrm{d}}$ should prevent valley mixing and lead to helicity control with longer valley lifetime, desirable for spin-valleytronics. 

\begin{figure}
\includegraphics[scale=0.51]{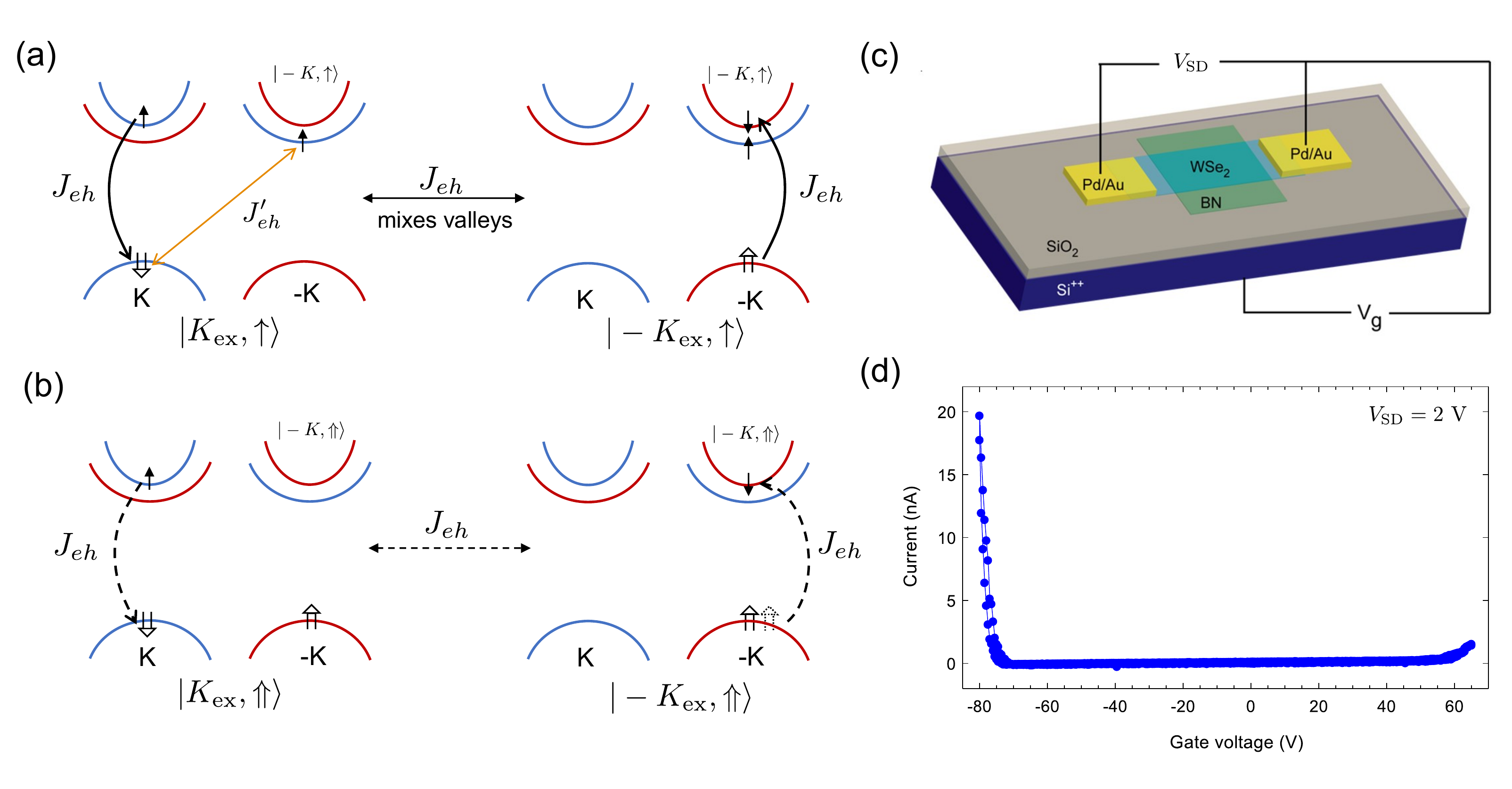}
\caption{{\bf Singly charged QDs and charge control in monolayer WSe$_2$ FET.}  {\bf a,} Schematic illustration of the single-particle states comprising a negatively charged QD ($X^-_{\mathrm{d}}$) with exciton in the K valley and the excess electron in the -K valley (left panel). Electron and hole spins are depicted with single- and double-line arrows, respectively. Electron spin-up (-down) bands appear in blue (red) color. Long range exchange interaction $J_{eh}$ (electron and hole in the same valley) mixes the opposite valley and same valley $X^-_{\mathrm{d}}$ (right panel) configurations. Consequently, the valleys of the exciton in $X^-_{\mathrm{d}}$ are mixed, while the spins of the excess electron are not. The short range exchange ($J'_{eh}$) splits the same valley and opposite valley configurations {\bf b,} Schematic illustration of the single-particle states comprising a positively charged QD ($X^+_{\mathrm{d}}$) with exciton in the K valley and the excess hole in the -K valley (left panel). $J'_{eh}$ is quenched due to holes forming a singlet and $J_{eh}$ is expected to be strongly suppressed because of Pauli blocking when charges are localized in a QD (right panel). {\bf c, d}  Monolayer WSe$_2$ field effect transistor (FET) device. (c) Schematic of the FET device. (d) FET current {\textit I$_{\mathrm{SD}}$} versus gate voltage curve measured with a source-drain voltage,  {\textit V$_{\mathrm{SD}}$}  = 2 V.}
\end{figure}

As shown in Fig.~1c, we employ a monolayer WSe$_2$ gate-controlled, charge tunable FET device, to obtain charged QDs (see Methods). By electrostatic doping, we inject free holes or electrons in the monolayer sample which then contribute to a current under an applied source-drain bias. Fig.~1d show that the sample has a higher propensity for hole-doping than electron as can be seen from the dominant hole current at negative gate voltage (V$_\mathrm{g}$). These free carriers can then be trapped in QDs, giving rise to localized charges in QDs. In our experiments, we operate close to the hole-doped regime of V$_\mathrm{g}$ where only free holes are expected to be present in the monolayer. Our device should be contrasted with a tunnel-coupling device where charge carriers tunnel in and out of the QD from nearby leads resulting in charge-controlled emission~\cite{HoegelePRL2004}.

We perform gate voltage (V$_\mathrm{g}$)-dependent photoluminescence (PL) spectroscopy on the monolayer WSe$_2$ FET device at low incident powers (see Methods). Fig.~2a shows a set of peaks appearing at a certain negative V$_\mathrm{g}$ when the sample is expected to be lightly hole-doped. For example, at V$_\mathrm{g}$ $\sim$ -10 V,  a single peak labeled S1 and a doublet labeled D1 appear simultaneously and spectrally wander in an identical manner, as highlighted by solid circles (Fig.~2a, left panel). Thus, we can conclude that the peaks S1 and D1 arise from the same QD. Likewise, at another location on the sample, S2/D2 group (see Supplementary Information for S6/D6 group) also display the same turn-on voltage and jittering pattern (Fig.~2a). We notice that the energy splitting of the doublets is $\sim$ 600 $\mu$eV which is consistent with the fine structure splitting resulting from the anisotropic e-h exchange interaction reported in previous studies on optically active neutral QDs in TMDs~\cite{SrivastavaNNano2015,KoperskiNNano2015,HeNNano2015}.  Thus, we assign D-peaks to neutral QDs, $X^0_{\mathrm{d}}$. 

\begin{figure}
\includegraphics[scale=0.53]{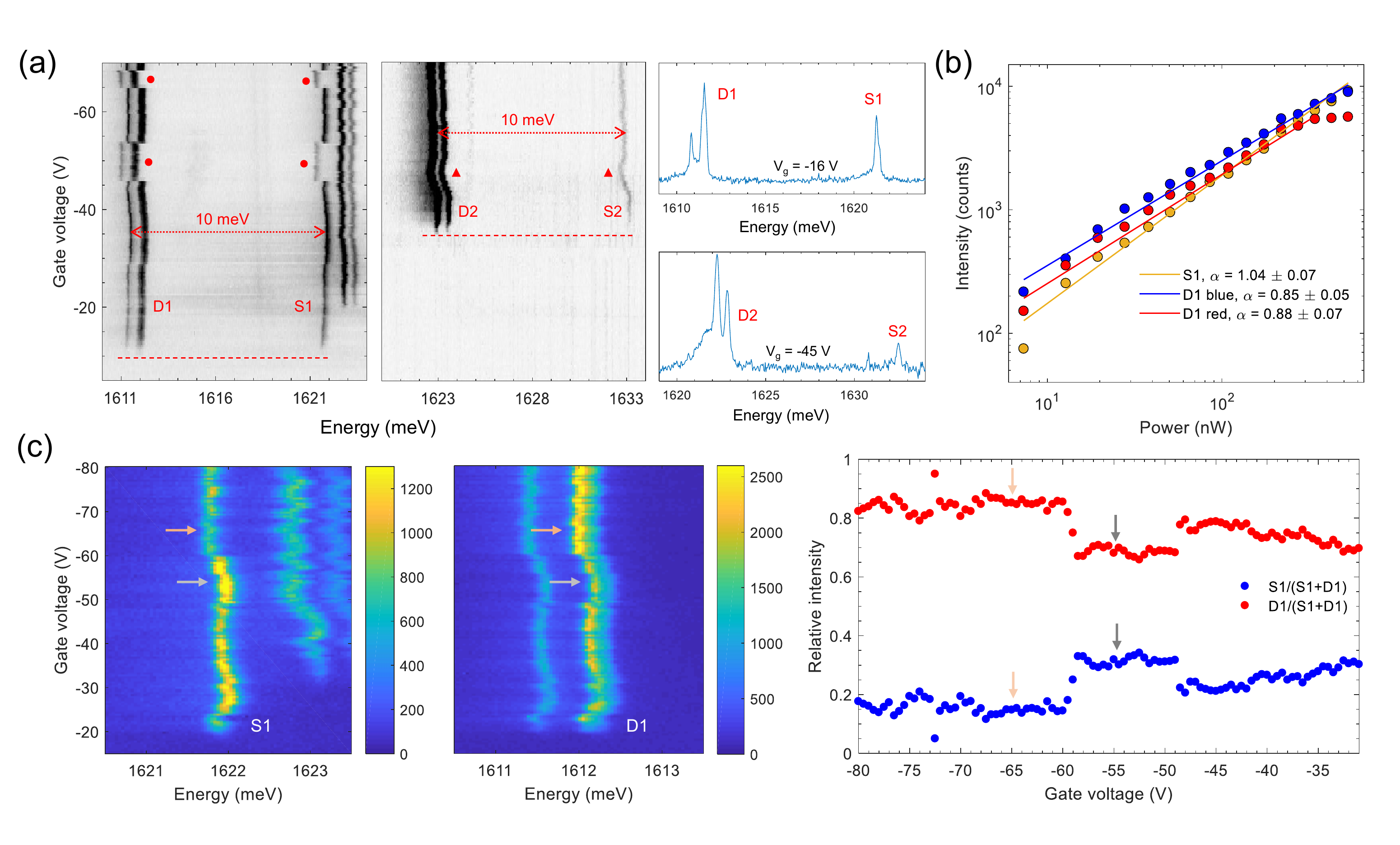}
\caption{{\bf Charged and neutral QDs in monolayer WSe$_2$.} {\bf a,} Left: PL intensity maps as a function of gate voltage ($ V_\mathrm{g}$). S1 and D1, as well as S2 and D2,  have correlated spectral jittering pattern (highlighted by solid symbols) and the same turn-on voltage (dash line), respectively. As a result, they are assigned to the same QD groups. The energy spacing between the S- and D-peaks is $\sim$10 meV for all QDs, Right: Cross-sectional PL spectra at a fixed $V_\mathrm{g}$ showing clear single and double peak features. {\bf b,} Power dependence of the S1-D1 group. The lines are power-law fitting $I \propto P^\alpha$. The extracted values of $\alpha$ are similar for S1 and D1 peaks. {\bf c,} Anti-correlated intensity between S1 and D1. Left: A $ V_\mathrm{g}$-dependent PL intensity map of S1-D1 group. S1 is stronger (weaker) when D1 is weaker (stronger), as indicated by gray (light orange) arrows. Right: Extracted relative intensity of S1 (blue dots) and D1 (red dots).  Excitation wavelength, $\lambda$, for S1-D1 and S2-D2 groups are 735~nm and 747~nm, respectively.}
\end{figure}

Next, we notice that the energy spacing between in S- and D-peaks is $\sim$ 10 meV for all QDs, with S-peaks located at higher energy. Unlike neutral and evenly-charged excitons where e-h exchange interaction causes a fine structure splitting, the e-h exchange interaction in a singly, positively charged QD is expected to vanish, as discussed earlier. As the sample is devoid of electrons in the range of V$_\mathrm{g}$ where S-peaks are observed, we assign S-peaks to $X^+_{\mathrm{d}}$. We contrast our findings with recently reported negatively charged QDs in WSe$_2$ where a fine structure splitting was seen, unlike our case of $X^+_{\mathrm{d}}$~\cite{ChakrabortyNanoLett2018}. We note that  $X^+_{\mathrm{d}}$ state has a binding energy ($E_{X^0_{\mathrm{d}}}$ $-$ $E_{X^+_{\mathrm{d}}}$) of -10 meV with respect to $X^0_{\mathrm{d}}$, which could originate from the Coulomb repulsion between holes and details of electron and hole wave functions in the QD. Indeed, negative binding energy of $X^+_{\mathrm{d}}$ has been reported in InGaAs QDs as well, as opposed to $X^-_{\mathrm{d}}$, and attributed to the different confining potential for electrons and holes~\cite{RegelmanPRB2001}. Fig.~2b shows that the excitation power-dependence of peaks in the S1-D1 group exhibits a similar power-law behavior consistent with our assignments and rules out other possible origins of S-peaks, such as a positively charged biexciton (see also Supplementary Information).

The coexistence of neutral QDs, $X^0_{\mathrm{d}}$, and positively charged QDs, $X^+_{\mathrm{d}}$, at V$_\mathrm{g}$ $<$ 0 during integration times on the order of tens of seconds indicates that the excess hole is trapped and released by $X^0_{\mathrm{d}}$ on a much faster timescale. However, as shown in Fig.~2c, occasionally an anti-correlation is seen in the intensity of S- and D-peaks at longer timescales. As indicated by arrows, S1 is weaker (stronger) when the intensity of D1 increases (decreases). This anti-correlation in intensities of S1- and D1-peaks is consistent with the picture that the excess hole is captured (released) by the QD during emission at the energy of S- (D-) peaks and also corroborates our claim that the peaks arise from the same QD (see also Supplementary Information). 

\begin{figure}
\includegraphics[scale=0.55]{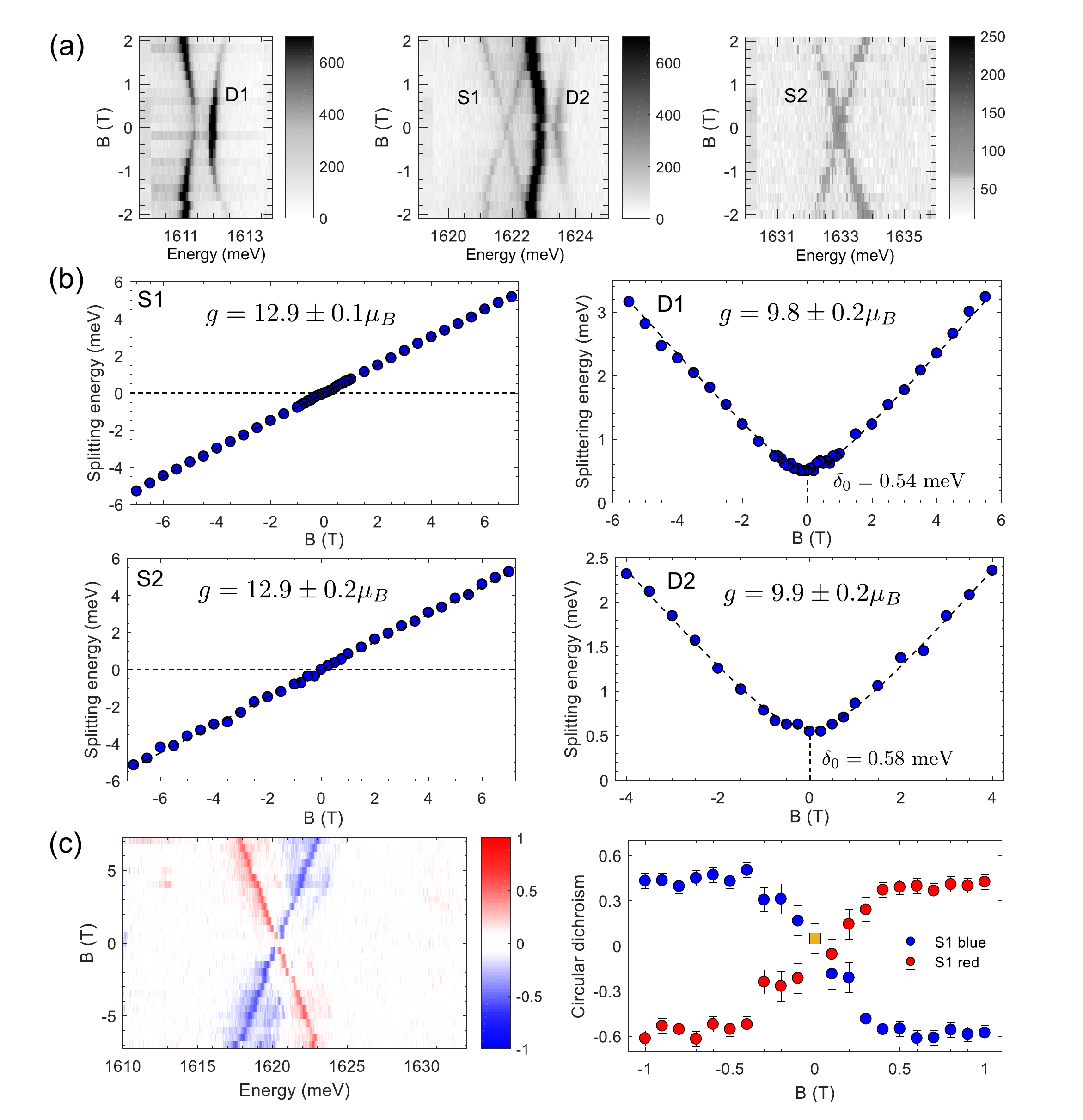}
\caption{{\bf Zeeman splitting and circular dichroism (CD) of a positively charged QD $X^+_{\mathrm{d}}$ in monolayer WSe$_2$.} {\bf a,} PL intensity map as function of magnetic field ($B$) (from -2 to 2~T). Both singlets (S1 and S2) and doublets (D1 and D2) show Zeeman splitting. Unlike the D-peak, the S-peak shows a linear, `X'-shaped splitting. Excitation wavelength, $\lambda$ = 735~nm. {\bf b,} g-factors of S1, D1, S2 and D2. D1 and D2 have the same g-factor of $\sim$10 and zero field splitting energy ($\delta_0$) of $\sim$0.56~meV, consistent with localized, neutral excitons $X^0_{\mathrm{d}}$. S1 and S2 display larger g-factor of $\sim$13 and no splitting at zero $B$ field (dash lines in the plot indicate $\delta_0$ = 0~meV). The values of g-factor are shown only in magnitude.  Excitation wavelength, $\lambda$ = 747~nm. {\bf c,} Left: CD, ($I_{\sigma^+}$ - $I_{\sigma^-}$)$/$($I_{\sigma^+}$ + $I_{\sigma^-}$), of S1 in $B$ (from -7 to 7~T with step of 0.5~T), where $I_{\sigma^+}$ ($I_{\sigma^-}$) denotes the intensity of the $\sigma^+$ ($\sigma^-$) circularly polarized emission. S1 does not have circular component at zero $B$, but the magnitude of CD increases with increasing $B$ field. Right: Scatter CD plot of S1 at low $B$ field (from -1 to 1~T with step of 0.1~T). S1 blue (red) peak is shown in blue (red) dot. The CD of S1 singlet at  $B$ = 0~T is depicted with yellow square. Sizable circular components are recovered even at a small $B$. At $B$ = -0.2 T, CD of  S1 red (blue) peak reaches -27$\%$ (31$\%$). Excitation laser is linearly polarized, $\lambda$ = 750~nm. }
\end{figure}

Having established that we observe positively charged and neutral excitons from the same QD, we perform polarization-resolved magnetic field ($B$) measurements in Faraday configuration ($B$ perpendicular to the sample). Fig.~3a shows that both S- and D-peaks display a Zeeman splitting in $B$ (see also Supplementary Information). However, unlike the D-peak, the S-peak shows a linear, `X'-shaped splitting consistent with the behavior of singly charged QD with no fine structure splitting~\cite{HoegelePRL2004}. We extract the corresponding g-factors which are plotted in Fig.~3b. The g-factor ($\sim$ 10) of D1 and D2 is consistent with previous studies of neutral QDs in WSe$_2$~\cite{SrivastavaNPhys2015,MacNeillPRL2015,LiPRL2014,StierNComm2016}.  A larger g-factor ($\sim$ 13) is observed for S-peaks and is consistent with the trend that $X^{\pm}$ (free trions) have a larger g-factor than $X^0$ (free exciton)~\cite{SrivastavaNPhys2015}. This difference in g-factors arises from the Coulomb interactions between electron and holes in the charged exciton state i.e., the initial state of the optical recombination process.  

Next, we analyze the polarization of S1 peak as function of $B$. Fig.~3c shows circular dichroism (CD) for linearly polarized excitation laser where CD is defined as ($I_{\sigma^+}$ - $I_{\sigma^-}$)$/$($I_{\sigma^+}$ + $I_{\sigma^-}$), with $I_{\sigma^+}$ ($I_{\sigma^-}$) denoting the intensity of the $\sigma^+$ ($\sigma^-$) circularly polarized emission. The linearly polarized excitation laser is $\sim$ 33 meV blue-detuned with respect to the emission at $B$ = 0. At $B$ = 0, we find that CD is vanishingly small, implying S1 is unpolarized (see Supplementary Information for linear basis measurements). As $X^+_{\mathrm{d}}$ is doubly degenerate at $B$ = 0 with energies of $| K_{\mathrm{ex}}, \Uparrow \rangle$ and $| -K_{\mathrm{ex}}, \Downarrow \rangle$ being equal, an unpolarized emission indeed is expected if the excitation laser is not exactly resonant with $X^+_{\mathrm{d}}$. As the $B$-field is increased, CD increases in magnitude implying that the split peaks become circularly polarized. The $\sigma^+$ ($\sigma^-$) emission is at lower (higher) energy at positive $B$, and shifts to higher (lower) energy at negative $B$. This can be understood by the fact that in finite $B$, the degeneracy of $K_{\mathrm{ex}}$ and $-K_{\mathrm{ex}}$ is lifted due to valley Zeeman effect~\cite{SrivastavaNPhys2015,AivazianNPhys2015}. As the emission from $X^+_{\mathrm{d}}$ takes place with $K_{\mathrm{ex}}$ ($-K_{\mathrm{ex}}$) recombining to give $\sigma^+$ ($\sigma^-$) polarized photon, we expect circularly polarized emission in finite $B$. We note that S1 gets a sizable circular component even at a small $B$ ($B$ $\sim$ 0.3 T) as there is no exchange interaction to overcome unlike in the case of $X^0_{\mathrm{d}}$. The applied $B$ does not influence the circular polarization of S1 once CD is saturated.

\begin{figure}
\includegraphics[scale=0.5]{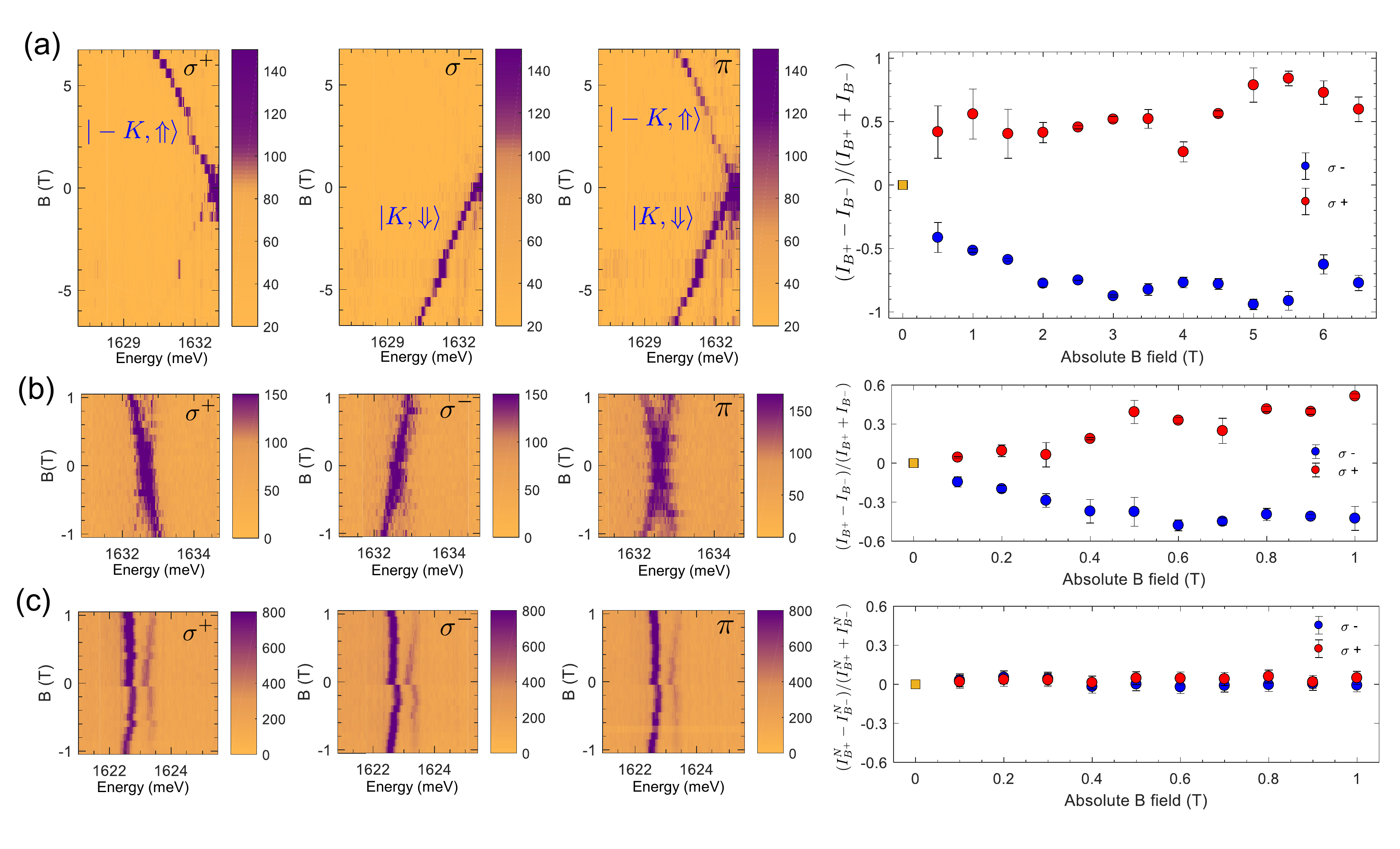}
\caption{{\bf Optical control of singlet emission in $B$ field.} {\bf a,} Left: PL intensity map of S2 red peak as a function of $B$ field with $\sigma^-$, $\sigma^+$ and linear ($\pi$) excitation (from -6.5 to +6.5~T with step of 0.5~T).  The $\sigma^-$($\sigma^+$) polarized emission corresponds to recombination of -K$_{\mathrm{ex}}$ (K$_{\mathrm{ex}}$) with $\sigma^-$ ($\sigma^+$) polarization. S2 red peak is $\sigma^-$ and $\sigma^+$ polarized at negative and positive $B$ field, respectively. S2 red peak is intense when incident laser is co-polarized with respect to emission, and disappears when it is cross-polarized. Linear excitation, which can be considered as the superposition of  $\sigma^+$ and $\sigma^-$ polarized states, does not have selective control of emission preference unlike circular excitation. Right: Extracted $B$ dependent ratio, ($I_{B^+}$ - $I_{B^-}$) $/$ ($I_{B^+}$ + $I_{B^-}$), of S2 red peak, where $I_{B^+}$($I_{B^-}$) denotes to the intensity measured under positive (negative) $B$.  The ratio approaches unity at high $B$, implying almost total optical control. {\bf b,} Left: PL color plot of S2 peak as a function of $B$ with $\sigma^-$, $\sigma^+$ and linear ($\pi$) excitation (from -1to +1~T with step of 0.1~T). Right: Extracted ratio ($I_{B^+}$ - $I_{B^-}$) $/$ ($I_{B^+}$ + $I_{B^-}$) of S2 red peak. Excitation wavelength, $\lambda$ = 747~nm in (a, b). {\bf c,} Left: $B$ dependent PL color plots of D2 doublet from -1 T to +1 T with similar detuning energy as to S2 in (a, b). $\lambda$ = 752~nm. No observable selectively in emission is seen in D2 upon changing the polarization of excitation. Right: $B$ dependent ratio, ($I^N_{B^+}$ - $I^N_{B^-}$) $/$ ($I^N_{B^+}$ + $I^N_{B^-}$), where $I^N_{B^+}$ ($I^N_{B^-}$) is the intensity of D2 red peak normalized by the sum intensity of D2 (red+blue). There is almost no difference between $\sigma^-$ and $\sigma^+$ excitation, implying that there is no observable optical control on the neutral exciton $X^0_{\mathrm{d}}$.}
\end{figure}

As shown above, the degeneracy of $X^+_{\mathrm{d}}$ is lifted in non-zero $B$, which allows us to spectrally distinguish between the spin-valley states ($|-K,\Uparrow\rangle$ or $|K,\Downarrow\rangle$) of the excess hole. As the valley index of the exciton in $X^+_{\mathrm{d}}$ is not mixed, we expect that controlling the helicity of the excitation laser should result in selective initialization of spin-valley of excess hole in $X^+_{\mathrm{d}}$. Fig.~4a shows the $B$ dependence of S2 (low energy peak) for $\sigma^+$, $\sigma^-$ and linear ($\pi$) excitation.
Indeed, by using $\sigma^-$ excitation, one only observes the S2 red peak at negative $B$ which corresponds to recombination of $-K_{\mathrm{ex}}$ with $\sigma^-$ polarization (co-polarized with excitation) while the emission from $K_{\mathrm{ex}}$ with $\sigma^+$ polarization (cross-polarized with excitation) completely disappears. The opposite case is observed with $\sigma^+$ excitation  while for linear excitation both branches are observed (see also Supplementary Information). As the valley index of the excess hole is opposite to that of the recombining exciton, we conclude that under $\sigma^-$ ($\sigma^+$) excitation, we can control the spin-valley state of the excess hole to be $| K, \Downarrow\rangle$ ($| -K, \Uparrow\rangle$). Remarkably, this helicity control is present even at smallest $B$-field as long as the splitting of the two peaks can be resolved (Fig.~4b). We quantify the strength of this control by calculating the ratio, ($I_{B^+}$ - $I_{B^-}$) $/$ ($I_{B^+}$ + $I_{B^-}$), where $I_{B^+}$ ($I_{B^-}$) is the intensity measured under positive (negative) $B$ (Fig.~4a-b, right panel). This ratio which is about 50 \% at 0.5 T and approaches unity at higher fields implying almost total control. We expect an even higher control at small $B$ as the detuning to  $X^+_{\mathrm{d}}$ is further reduced. Our observations suggests that by controlling helicity of the excitation, we selectively excite $K_{\mathrm{ex}}$ or $-K_{\mathrm{ex}}$ of $X^+_{\mathrm{d}}$ which in turn fixes the excess hole to be from the opposite valley, leading to its initialization in a known spin-valley state. As long as the exciton of $X^+_{\mathrm{d}}$ survives, the spin-valley of the excess hole maintains its state. Even after the recombination of exciton, which takes place after a few nanoseconds~\cite{SrivastavaNNano2015,KoperskiNNano2015,HeNNano2015}, one expects the spin-valley of single, localized hole to be preserved for a much longer time. In fact from the valley lifetime of free holes~\cite{JiangNComm2018,DeyPRL2017,KimScience2017}, we expect a single spin-valley lifetime on the order of microseconds, if not longer. On longer timescales, valley relaxation could be mediated by hyperfine interaction with nuclear spins, which is expected to be quite small in TMDs~\cite{WuPRB2016,SharmaPRB2017}. 
 
Although this observation seems very similar to valley polarization of $X^0$ in TMDs, there is a crucial difference. The reason for valley polarization in PL of $X^0$, even in presence of e-h exchange, is that the PL lifetime ($\sim$ ps) is faster than (or comparable to) the valley mixing time of $\sim$ 4 ps assuming an exchange energy of $\sim$ 1 meV~\cite{YuNComm2014}. On the other hand, the PL lifetime of QDs is on the order of nanoseconds and the helicity control of spin-valley stems from quenching of e-h exchange in $X^+_{\mathrm{d}}$. To further check this claim, we perform similar measurements on D2 ($X^0_{\mathrm{d}}$) with similar detuning energy. D-peaks are linearly polarized at $B$ = 0 and become circularly polarized at higher $B$~\cite{Schwarz2DM2016} (see Supplementary Information). Fig~4c shows that there is almost negligible helicity control of the circularly polarized branches of D2 even under $B$, suggesting fast valley relaxation due to e-h exchange (see also Supplementary Information). 

In conclusion, we have observed positively charged $X^+_{\mathrm{d}}$ and neutral $X^0_{\mathrm{d}}$ states of the same QD in a monolayer WSe$_2$ FET device.  The charged QD hosts an excess hole with a net spin-valley which is opposite to the valley of the e-h excitation. We find that e-h exchange interaction responsible for valley mixing is quenched in positive, singly charged QDs which enables helicity control of its spin-valley under small magnetic fields. Our results show that spin-valley degree is robust in optically active TMD QDs and enables valleytronics on single localized charge carriers. 
 \\

\textbf{Methods}
\\
\textbf{Sample fabrication.} We use polydimethylsiloxane (PDMS) based dry transfer method to fabricate WSe$_2/$BN (crystals from HQ graphene) stack on a degenerately doped Si (Si$^{++}$) substrate with 285~nm SiO$_2$ on top. Electron beam lithography is used to deposit 30~nm Pd$/$80~nm Au metal contacts on WSe$_2$, which act as source and drain electrodes. The charge carrier density in WSe$_2$ is controlled by applying voltage (Keithley 2400 sourcemeter) to the Si$^{++}$ substrate, with the 285~nm SiO$_2$ acting as the gate dielectric.  
\\
\textbf{PL spectroscopy.} The sample is loaded into a closed-cycle cryostat (BlueFors cryogenics) equipped with magnetic field ranging from -8 to +8~T and cooled down to $\sim$ 3.5~K. A piezo controller (Attocube systems) is used to position the sample. Photoluminescence spectroscopy was performed using a home-built confocal microscope set-up. The emission was collected using an aspheric lens (0.55 NA) and directed to a high-resolution (focal length$:$ 750 mm) spectrometer where it was dispersed by a 1200 g$/$mm grating (blazed at 750~nm). A liquid nitrogen-cooled charge coupled device (Princeton Instruments SP-2750, PyLoN 1340 $\times$ 400 pixels CCD) was used as detector. We use a mode-hop-free tunable continuous-wave Ti:Sapphire laser (M Squared Lasers) with resolution of 0.1~pm and power of 300 or 400 ~nW (except during power dependence measurement) as excitation source. The spot size for the Ti:Sapphire laser  is $\sim$1~$\mu$m. Polarization of the incident laser is controlled by using a polarizer together with a full-wave liquid crystal retarder. Circular polarization-resolved measurements were performed with a $\lambda/$4 (achromatic, 690 - 1200 nm) plate placed before the Wollaston prism. One can collect $\sigma^+$ and $\sigma^-$ components of the polarization simultaneously. The influence of blinking on polarization measurements is eliminated in this setup. Circular emission from QD is converted into linearly-polarized light after passing through the $\lambda/$4 plate. Wollaston prism separates light into $s$- and $p$-components. Another $\lambda/$4 plate (zero order @ 780~nm) is placed after the Wollaston prism to convert the linearly polarized light into circularly polarized signal, so that the signal will be insensitive to the grating efficiency. In all the magnetic field dependence measurements, $B$ is applied perpendicular to the plane of the sample.
\\

\vspace{1 cm}

\textbf{Acknowledgments} We acknowledge many enlightening discussions with Ata\c{c} Imamo\u{g}lu and Martin Kroner. We also acknowledge technical help from Timothy Neal and Eva Liu. A. S. acknowledges support from Emory University startup funds and NSF through the EFRI program-grant \# EFMA-1741691. Q.X. gratefully acknowledges strong support from Singapore National Research Foundation via NRF-ANR joint grant (NRF2017-NRF-ANR002 2D-Chiral) and Singapore Ministry of Education via AcRF Tier2 grant (MOE2017-T2-1-040) and Tier1 grants (RG 113/16 and RG 194/17)\\

\textbf{Author Contributions} X. L., X. C., S. D. and Q. Y. carried out the quantum dot measurements. X. L. and X. W. prepared the samples. A. S. and Q. X. supervised the project. All authors were involved in analysis of the experimental data and contributed extensively to this work.\\

\textbf{Author Information} The authors declare that they have no
competing financial interests. Correspondence and requests for
materials should be addressed to A.S. (ajit.srivastava@emory.edu) or X.L. (xin.lu2@emory.edu).

\textbf{Competing financial interests} The authors declare no competing financial interests.

\newpage

\end{document}